\documentclass[aps,pre,reprint]{revtex4-1}
\usepackage[T1]{fontenc}
\usepackage{amsmath}
\usepackage[latin1]{inputenc}
\usepackage{graphicx}

\usepackage{epsfig}

\usepackage{color}

\begin{document}
\title{Reentrant Network Formation in Patchy Colloidal Mixtures under Gravity}
\begin{abstract}
We study a two-dimensional binary mixture of patchy colloids in sedimentation-diffusion-equilibrium using Monte Carlo
simulation and Wertheim's theory. By tuning the buoyant masses of the colloids we can control the 
gravity-induced sequence of fluid stacks of differing density and percolation properties. We find complex stacking sequences with
up to four layers and reentrant network formation, consistently in simulations and theoretically using only the bulk phase diagram as input.
Our theory applies to general patchy colloidal mixtures and is relevant to understanding experiments under gravity.
\end{abstract}

\author{Daniel de las Heras}
\affiliation{Theoretische Physik II, Physikalisches Institut,
  Universit{\"a}t Bayreuth, D-95440 Bayreuth, Germany}
\author{Lucas L. Treffenst{\"a}dt}
\affiliation{Theoretische Physik II, Physikalisches Institut,
  Universit{\"a}t Bayreuth, D-95440 Bayreuth, Germany}
\author{Matthias Schmidt}
\affiliation{Theoretische Physik II, Physikalisches Institut,
  Universit{\"a}t Bayreuth, D-95440 Bayreuth, Germany}

\maketitle

Patchy colloids~\cite{in1,in2} are nano- to micron-sized particles with a solid core and a discrete number of interaction sites or patches.
Pairs of patches can form reversible bonds which tie the particles together. By tuning the shape of the colloids, and the number and types of patches, one
can find complex phase behaviour and novel types of phases. Examples are empty liquids~\cite{emptyscio}, network fluids with 
pinched phase behaviour~\cite{PhysRevLett.106.085703}, lower critical points~\cite{PhysRevLett.111.168302}, and self-assembly into complex structures
~\cite{chen2011directed,0953-8984-25-19-193101,kraft2012surface}. The bulk phase behaviour of model patchy
colloids has been investigated via computer simulations, using molecular dynamics
and Monte Carlo, see e.g. Refs.~\cite{emptyscio,mcmcmc,doi:10.1021/jp056380y}. Wertheim's
association theory~\cite{wertheim} is a widely used approach since it accurately predicts~\cite{wttavares,wtmarshall,mcwt} 
the structures observed in computer simulations. We hence have appropriate tools to analyse the bulk behaviour of model patchy colloids.
However, in a real experimental realization~\cite{emptyexp}, it is often unavoidable that colloids are subject to gravity.

Gravitational and thermal energies are typically comparable in colloidal systems. Hence, gravity can have a strong effect and
drive new complex phenomena, especially in mixtures, where colloidal particles can float on top of a fluid of lighter colloids~\cite{Piazza}, and
very complex sequences of stacks can occur even for systems with rather generic bulk phase behaviour~\cite{multiphase1,multiphase2,floating}.
Relating the bulk phase diagram to sedimentation-diffusion-equilibrium is therefore of vital importance to
adequately comparing theoretical studies of bulk behaviour to findings in sedimentation experiments~\cite{JeR}.

Here, we study theoretically the bulk phase behaviour of a two-dimensional mixture of patchy colloids, and use it to obtain the sedimentation behaviour
via a local density approximation~\cite{stacking1}. We compare the results to extensive Monte Carlo simulations of the same mixture under gravity.
Despite the simplicity of the bulk phase diagram of the model, we find very rich phenomenology under gravity, including reentrant 
network formation. Moreover, we show how to use the bulk phase diagram to systematically predict the sedimentation behaviour.

\begin{figure}
\includegraphics[width=0.75\columnwidth]{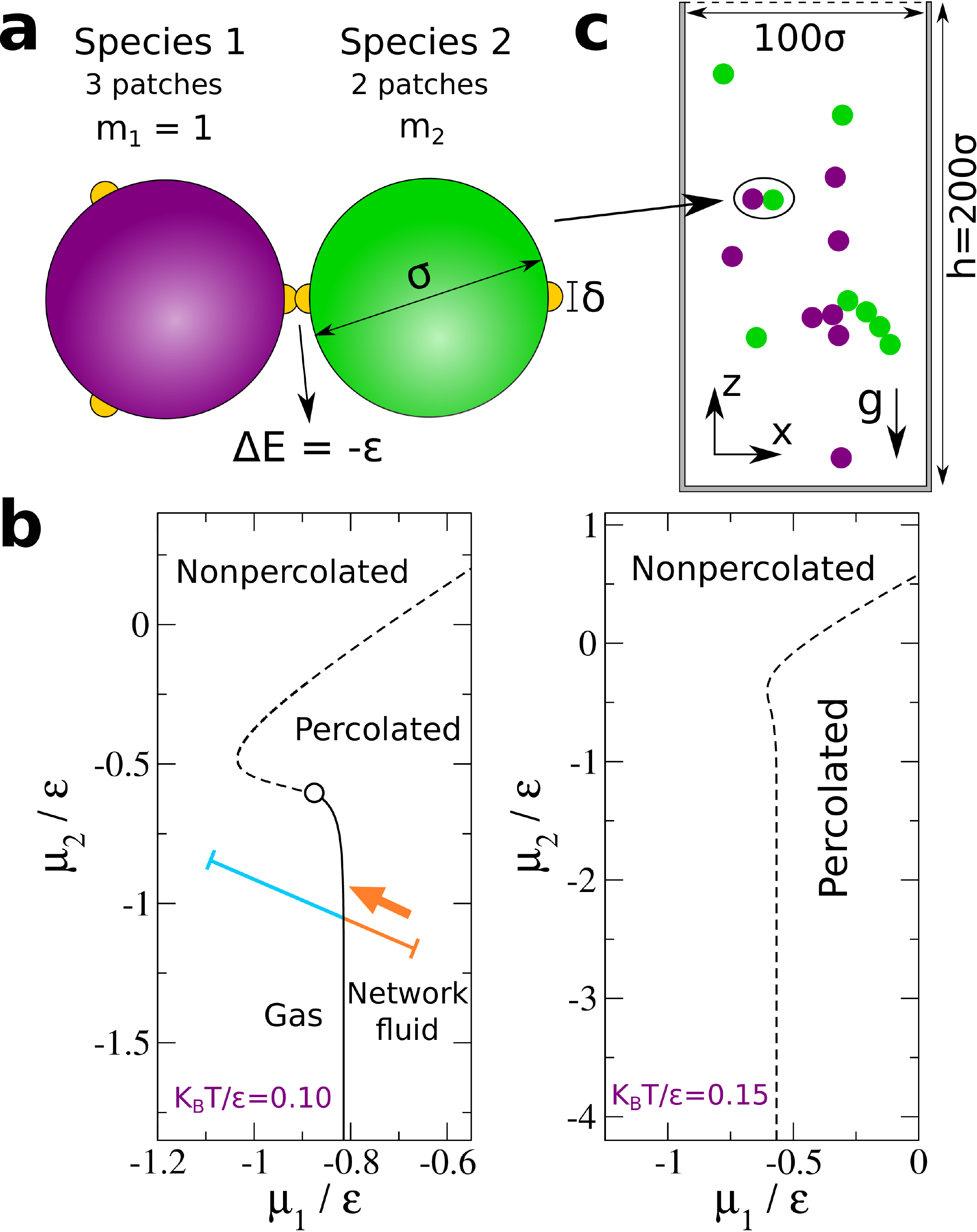}
\caption{(color online) (a) Illustration of the two types of patchy particles showing the inner hard core
(diameter $\sigma$) and the patches (diameter $\delta$). (b) Bulk phase diagram of the mixture at
$k_\text{B}T/\epsilon=0.10$ (left) and $0.15$ (right) in the $\mu_1-\mu_2$ plane of chemical potentials 
according to Wertheim's theory. The black solid line is the binodal. The dashed line is the percolation
line. The empty circle is the critical point. The binodal and the percolation line divide the phase
diagram into percolated and nonpercolated regions as indicated.
The straight line is an example of a sedimentation path. 
The arrow indicates the direction of the path from bottom to top. (c) Geometry of the simulation box to
study sedimentation: a rectangular vessel (width $100\sigma$ and height $h=200\sigma$) made of hard walls.} 
\label{fig1}
\end{figure}

Colloidal particles vary in size from nano- to micro-meters. Here, we model the colloids as a 
hard core (hard disk of diameter $\sigma$) with interaction sites or patches on the surface,
see Fig.~\ref{fig1}a. The patches are disks of diameter $\delta=(\sqrt{5-2\sqrt3}-1)\sigma/2\approx0.12\sigma$ with their
centers of mass equispaced on the surface of the particle. Two patches form a bond if they overlap, in which case the
internal energy of the system decreases by $\epsilon=1$. The chosen size of the patches ensures that only bonds
involving two patches can be formed~\cite{sciortino2d}. The two species differ in their number of patches and their buoyant
masses. Species $1$ ($2$) has three (two) patches, and we consider a range of ratios of the buoyant masses. 

In Fig.~\ref{fig1}b we show the bulk phase diagram in the plane of chemical potentials of
the two species, $\mu_1$ and $\mu_2$,
according to Wertheim's theory for scaled temperatures $k_\text{B}T/\epsilon=0.10$ and $0.15$, 
with $k_\text{B}$ the Boltzmann constant. We use the implementation of the theory
described in~\cite{2dh} with the bonding volume corresponding to our type of patches, $v_b=4.4\cdot10^{-3}\sigma^2$. 
Previous studies~\cite{sciortino2d,2dh} of this system have shown
theoretical results to be in very good agreement with Monte Carlo simulation data.
At the lower temperature there occurs a first order phase transition between a high- and a low-density fluid phase.
In the $\mu_1-\mu_2$ plane, the binodal, at which both fluid phases coexist, has a
vertical asymptote ($\mu_2\rightarrow-\infty$) that tends to the value of $\mu_1$ at the phase
transition of the monocomponent system with three patches (species $1$). A pure system with only two patches
can form only chains and does not undergo a phase transition due to the absence of branching.  Hence, the binodal of the
mixture ends at a critical point. We also plot the percolation line, which we have
obtained with the generalization of Flory-Stockmayer (FS) percolation theory~\cite{flory,stock} to binary
mixtures~\cite{herasNF}. FS theory is based on the assumption that the particles form tree-like clusters, i.e. closed loops are neglected,
which is consistent with Wertheim's theory. The mixture percolates if the probability that a patch is
bonded, $f_\text{b}$, is higher than a percolation threshold, $p$. For a mixture of bi- and
tri-functional particles~\cite{herasNF} 
\begin{equation}
p=\frac{x+2}{4x+2},
\end{equation}
where $x=N_1/(N_1+N_2)$  is the molar fraction of species $1$, with $N_i$ the number of
particles of species $i$. 
In the percolation region of the phase diagram, cf. Fig.~\ref{fig1}b, a transient cluster spans the entire system
volume. In the nonpercolated region the particles form clusters of only finite size. The
percolation line intercepts the binodal on the gas side, very close to the critical point.
Hence, the high-density phase is always percolated. We refer to this phase as a network fluid~\cite{herasNF}.
The second case, $k_\text{B}T/\epsilon=0.15$, is above the upper critical temperature of the system. Hence
the first-order gas-fluid transition is absent, but the percolation line divides the phase
space into percolated and nonpercolated regions.

We study theoretically and with Monte Carlo simulations the mixture under gravity, i.e.,
in sedimentation-diffusion-equilibrium. We simulate the system in the canonical ensemble,
fixing the number of particles $N_1$ and $N_2$, the system volume $V$,
and the temperature $T$. We confine the particles in a rectangular box of width $100\sigma$
and height $h=200\sigma$, see Fig.~\ref{fig1}c. The surface of the box consists of hard walls,
i.e., the hard cores of the colloids cannot penetrate the walls. We apply an external gravitational
field, $\phi_i(z)=m_igz$, where $g$ is the gravitational constant that we set to
$g=0.005\epsilon/(m_1\sigma)$,  $z$ is the vertical coordinate, and $m_i$ is the buoyant mass of
species $i$. We set $m_1=1$ as the unit of buoyant mass. To initialize the system we first run a
simulation at a high temperature and take the last configuration as the initial state of the
simulation at the temperature of interest. We run an equilibration stage until the energy fluctuates around a minimum 
value, and then perform $6\cdot10^7$ Monte Carlo steps to acquire the data. The data is 
averaged over three independent such simulations to improve statistics.

We compare the simulation results to theoretical predictions based on the concept of sedimentation
paths~\cite{stacking1,stacking2} in the bulk phase diagram. We extend this theory to the case
of systems exhibiting a percolation transition. We define a local chemical potential~\cite{stacking1,stacking2}:
\begin{equation}
\psi_i(z)=\mu_i^\text{b}-m_igz,\;\; i=1,2,\label{eq:localchepo}
\end{equation}
with $\mu_i^\text{b}$ the bulk chemical potentials in absence of gravity. Combining Eq. (\ref{eq:localchepo}) for $i=1,2$ we obtain the sedimentation path:
\begin{equation}
\psi_2(\psi_1)=a+s\psi_1\label{eq:path},
\end{equation}
where $a$ and $s$ are constants given by
\begin{eqnarray}
a & = & \mu_2^\text{b}-s\mu_1^\text{b},\nonumber\\
s & = & m_2/m_1\label{eq:constants}.
\end{eqnarray}
In general, the gravitational length of the colloids, $\xi_i=k_\text{B}T/(m_ig)$, is much larger than
any characteristic correlation length, which is typically of the order of the particles' sizes.
Hence, we can assume that locally, i.e., at each value of $z$, the state of the system is well described by an
equilibrium bulk state at the same chemical potential as the local chemical potential i.e.,
$\psi_i(z)=\mu_i$ for both species $i=1,2$. This local density approximation directly relates the sedimentation path,
Eq~(\ref{eq:path}), to the stacking sequence of the mixture. The sedimentation path is a
straight line segment in the plane of chemical potentials $\mu_1$ and $\mu_2$. Each crossing between the sedimentation path and a
boundary line between two phases corresponds to an interface in the sedimented sample. As an example, the path shown
in Fig.~\ref{fig1}b corresponds to a stacking sequence formed by a bottom network fluid and a top
gas, because the path crosses the binodal line once. 
\begin{figure*}
\includegraphics[width=1.4\columnwidth]{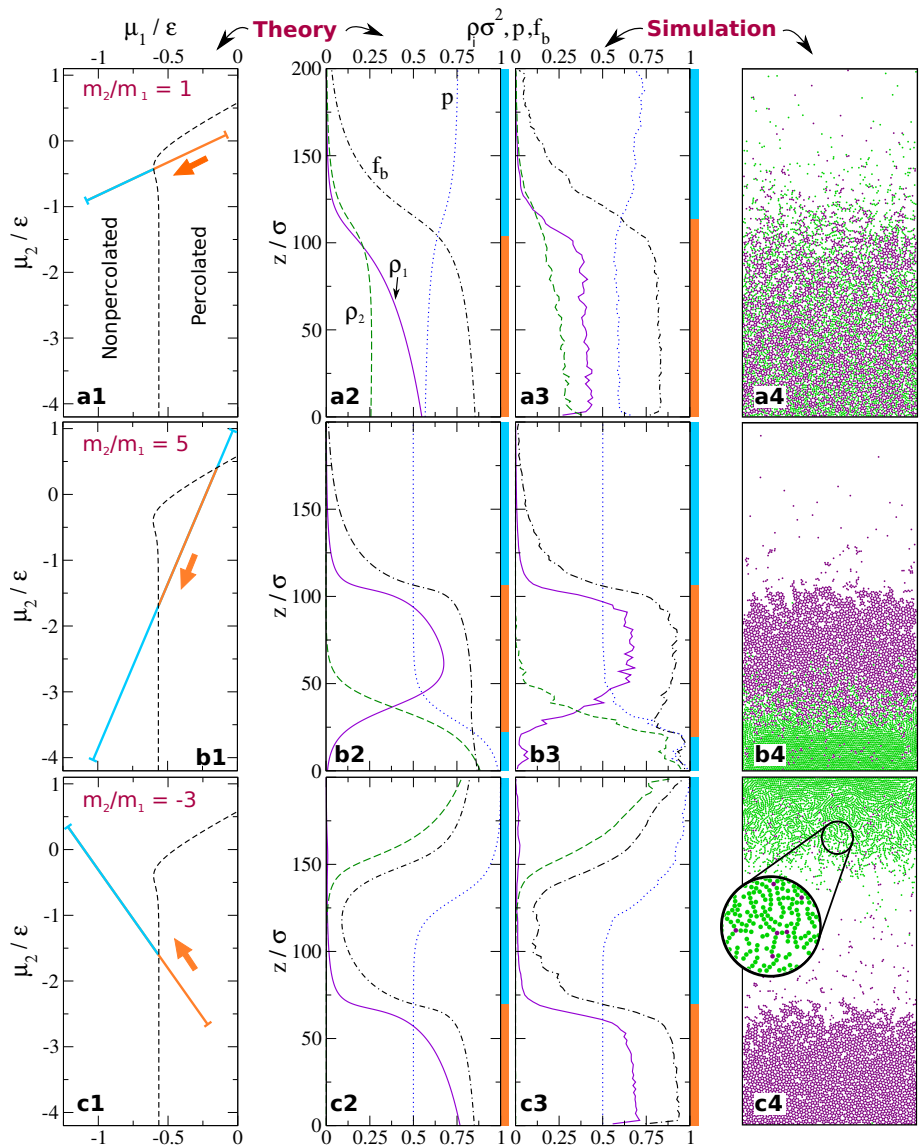}
\caption{(color online) Binary mixture of patchy colloids under gravity, $g=0.005\epsilon/(m_1\sigma)$.
(first column) Phase diagram in the plane of chemical potentials $\mu_1-\mu_2$ showing the sedimentation
path (straight line). The arrows indicate the direction of the path from the bottom to the top of
the sample. The path is colored in orange (cyan) in the percolated (nonpercolated) region.
The height of the sample is $h=200\sigma$. (second and third column) Density profiles $\rho_i$
of species $i=1$ (solid violet line) and species $i=2$ (dashed green line), percolation threshold $p$ (dotted
blue line), and probability that a patch is bonded $f_\text{b}$ (dot-dashed black line). The orange (cyan)
rectangle on the right of each panel indicates the percolated (nonpercolated) region, i.e., the region where $f_\text{b}>p$ ($f_\text{b}\le p$).
Results are according to theory (second column) and simulations (third column). The forth column shows
characteristic snapshots of the simulation. Each row correspond to a different ratio of buoyant
masses: $m_2/m_1=1$ (first row), $m_2/m_1=5$ (second row), and $m_2/m_1=-3$ (third row). The inset in (c4)
is a close view of a small region. In all cases the
temperature is $k_\text{B}T/\epsilon=0.15$, the composition of the mixture is $x=0.6$, and
in simulation $N=7500$.}
\label{fig2}
\end{figure*}
The length of the path in the $\mu_1-\mu_2$ plane
is proportional to the height of the container: $\Delta\mu_i=m_igh$ with $\Delta\mu_i$
the difference in chemical potential between the top and the bottom of the sample. The buoyant masses
fix the slope, Eq.~(\ref{eq:constants}), and the direction of the path, Eq.~(\ref{eq:localchepo}).

In order to compare to simulations, we calculate the average densities
\begin{equation}
\bar\rho_i=\frac{1}{h}\int_0^hdz\rho_i(z),\;\;i=1,2
\end{equation}
with $\rho_i(z)$ the number density profile of species $i$ which we obtain from the bulk equation of state
$\rho_i(\mu_1,\mu_2)$ by calculating the local chemical potentials at a given height. Then,
we move the path in the $\mu_1-\mu_2$ plane until the average densities equal those in the simulation, $\bar\rho_i=N_i/V$.

Fig.~\ref{fig2} shows representative results for three different
mixtures with different ratio of the buoyant masses. In all cases $N_1=4500$,
$N_2=3000$, and hence $x=0.6$. The scaled temperature is $k_\text{B}T/\epsilon=0.15$, so the percolation line is the only
feature present in the phase diagram.

First we analyse the case of equal masses $m_1=m_2$ (first row in Fig.~\ref{fig2}). The sedimentation path (a1) crosses
the percolation line once, which generates the sequence percolated-nonpercolated (from bottom to top). Theoretical and
simulation density profiles are shown in panels (a2) and (a3), respectively. We also show the bonding
probability $f_b$ and the percolation threshold $p$. Theory and simulation
results for all quantities agree quantitatively. Both show the same stacking sequence and the interface between
percolated and nonpercolated states, that occurs at $f_b=p$, is located at (almost) the same height.
The percolated phase is a homogeneous mixture of both species, i.e. there is no demixing. 
A snapshot of the simulation is shown in (a4). 

In the second row of Fig.~\ref{fig2} we increase the buoyant mass of bi-functional particles, $m_2=5m_1$.
The sedimentation path in the plane of chemical potentials 
(b1) is substantially longer and steeper than in the previous example. This allows the
path to cross the percolation line twice, generating a reentrant phenomenon. The path starts (bottom)
in the nonpercolated region at high chemical potentials. The corresponding layer is a high density
nonpercolated phase, which is rich in bi-functional particles. Then, the path crosses the percolation line, 
giving rise to a percolated layer. Close to the bottom interface the percolated layer
contains particles of both species. As the path enters the percolated region (increasing the
height) the state changes to a network mainly composed of tri-functional particles. Finally, the path
crosses the percolation line and a nonpercolated gas phase appears in the vessel. Again theory (b2) and simulation
results (b3) are in very good agreement with each other. The snapshot (b4) illustrates the system.

The third row of Fig.~\ref{fig2} shows results for $m_2=-3m_1$, hence
the mass density of species $2$ is smaller than the mass density of the solvent, and species
$2$ has a tendency to cream upwards. The path (c1),
now with negative slope, crosses the percolation line once. The stacking sequence is bottom percolated
and top nonpercolated, as in the case of equal masses (first row). However, in this case the bottom
percolated stack is almost a pure system of tri-functional particles, and the top nonpercolated phase
consists of two distinguishable regions. Close to the interface with the percolated region there is a gas phase and close to the top
of the sample there is a dense layer primarily composed of bi-functional particles. Both theory and simulations
show that this dense layer contains also a few particles with three patches, see inset in (c4), but not in sufficient number to induce
percolation.

\begin{figure}
\includegraphics[width=0.9\columnwidth]{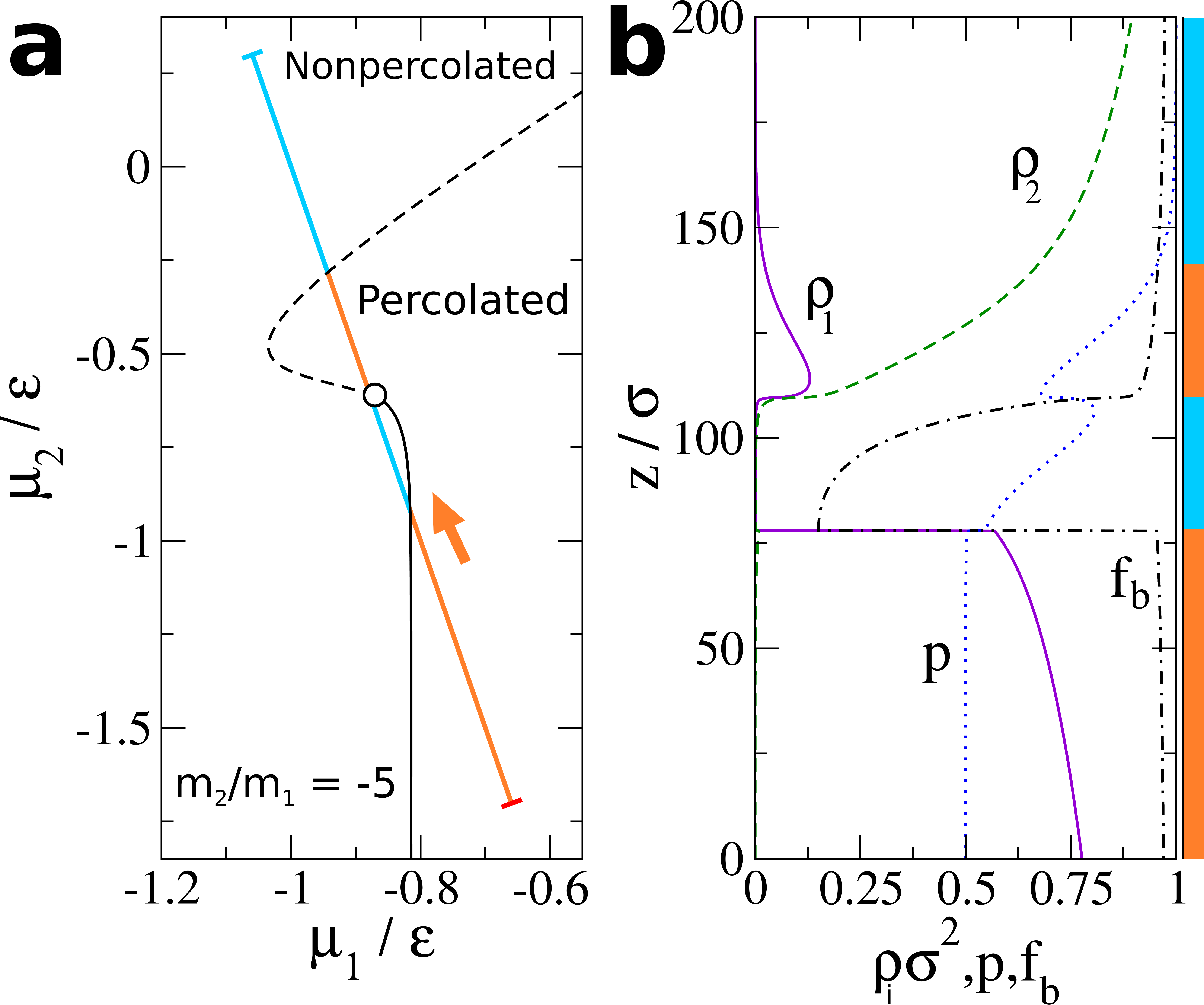}
\caption{(color online) (a) Bulk phase diagram according to Wertheim's theory in the plane of chemical potentials at $k_\text{B}T/\epsilon=0.1$
together with a sedimentation path (straight line). The path corresponds to the case $g=0.002\epsilon/(m_1\sigma)$,
$m_2=-5m_1$, $h=200\sigma$ and average densities $\bar\rho_1\sigma^{2}=0.288$, $\bar\rho_2\sigma^2=0.313$.
(b) Density profile $\rho_i$ of species $i=1$ (solid violet line) and species $i=2$ (dashed green line), percolation
threshold $p$ (dotted blue line), and bonding probability $f_\text{b}$ (dot-dashed black line) corresponding to
the sedimentation path shown in (a). The orange (cyan) rectangle on the right of panel (b) indicates the percolated (nonpercolated)
region, i.e., the region where $f_\text{b}>p$ ($f_\text{b}\le p$). The sedimentation path (a) has the same color-code.} 
\label{fig3}
\end{figure}

Given the S-shape of the percolation line, a straight sedimentation path can cross it three times, generating a reentrant phenomenon. 
This effect is very strong at low temperatures. In Fig.~\ref{fig3}a we show the phase
diagram at $k_\text{B}T/\epsilon=0.1$, together with a sedimentation path for the case $m_2=-5m_1$ and $g=0.002\epsilon/(m_1\sigma)$. The path
starts in the percolated region, then crosses the binodal entering the gas phase, and finally crosses twice the percolation line.
The corresponding sequence consists of four layers: bottom percolated-nonpercolated-percolated-nonpercolated. 
This sequence constitutes a reentrant phenomenon for both percolated and nonpercolated phases. The  
bottom percolated-nonpercolated interface is very sharp since the path crosses the binodal line in the bulk phase diagram,
see the density profiles in Fig.~\ref{fig3}b. The top nonpercolated-percolated interface is much wider. In this case,
the path crosses the percolation line instead of the binodal.
The upper percolated stack appears due to localized concentration of tri-functional particles, much heavier than the particles
with two patches. Hence, the gravitational energy would be lower if the heavier particles with three patches 
sedimented to the bottom stack. They are, however, far away from the bottom interface because they increase
the bonding probability of particles with two patches, and hence decrease the free energy of bonding. Hence it is the balance between
gravitational and bonding energies that drives the formation of this complex stacking sequence. 
Given the low temperature we have been unable to equilibrate the system with computer simulations. Advanced simulation techniques, such
as "virtual-move"~\cite{virtualmove} or "aggregation-volume-bias"~\cite{AVBMC} Monte Carlo might alleviate the equilibration problem.

The bulk phase diagram of the two-dimensional mixture analyzed here and its corresponding three-dimensional
mixture of bi- and tri-functional hard spheres~\cite{emptyscio,herasNF} are qualitatively the same. 
Hence, we are confident that the sedimentation scenario is similar in three dimensions, since only the topology of the 
bulk phase diagram is relevant for determining the behaviour under gravity.

Here we have used a strong gravitational field that affects the system at scales of 
a few particle lengths, as expected for micronsized colloids~\cite{Royall}. In the case
of nanoparticles, gravity affects the system at length scales of the order of $10^{4}-10^5$ particle sizes.
Our theory assumes the gravitational length to be much larger than any other correlation length, and hence it
is expected to perform even better for such systems. Our theory might not be accurate in the vicinity
of a critical point. However, we do not expect completely erroneous predictions, such as e.g. a different stacking sequence to occur. Nevertheless, the
influence of gravity on critical phenomena is an important research topic, see e.g.~\cite{B915788C},
 which remains to be investigated for the case of patchy colloids.

Our results demonstrate that the presence of lateral walls, used in simulations and certainly present in experiments,
does not have a negative impact on the development of complex stacking sequences.

Despite the simplicity of the bulk behaviour of our system, displaying only a single binodal and one percolation line, the sedimentation behaviour is quite complex.
We therefore expect a much richer phenomenology when analyzing sedimentation of patchy colloids with complex bulk phase diagram, such 
as {\it e.g.} those of patchy colloids with flexible bonds~\cite{complex11}. The ratio between buoyant masses is a key
parameter that partially controls the stacking sequence. This ratio can be experimentally tuned by 
fabricating colloids with inner cores~\cite{doi:10.1021/nl025598i} made of different materials or by changing the density of the solvent.

The theoretical
framework developed here is general and can be applied to study sedimentation of any patchy colloidal system using the bulk phase behaviour as input. 
Here, we have used Wertheim's theory but other theories or even simulation techniques can be used to provide the bulk behaviour.
We have shown how structural crossover, such as the percolation line, in the bulk phase diagram
relates to complex stacking sequences in sedimented systems. This extends our previous work~\cite{stacking1,stacking2}, which was aimed only at phase
coexistence. We have shown that the influence of gravity generates networks that are spatially inhomogeneous on the scale of the system
size.  This constitutes a further level of complexity of the self-assembly in patchy colloid systems. We have shown that
the type of stacking can be precisely controlled by tuning simple material properties, such as the buoyancy masses of the particles.
Our work paves the way for studying inhomogeneous patchy colloidal systems, where we expect a wealth of new phenomena occur, as is the case
in the fields of inhomogeneous liquids and simple colloidal dispersions.

DdlH acknowledges support from the Portuguese Foundation for Science and Technology (FCT) through project EXCL/fis-nan/0083/2012.
DdlH and LT contributed equally to this work.

\end{document}